\documentclass[%
 reprint,
superscriptaddress,
 amsmath,amssymb,
 aps,
pra,
]{revtex4-1}

\usepackage{amsthm,amssymb}
\usepackage{amsmath}
\usepackage{graphicx}
\usepackage{dcolumn}
\usepackage{bm}
\usepackage{multirow} 
\usepackage{enumerate}
\usepackage{algorithm}
\usepackage{algorithmic}

\theoremstyle{remark}

\newtheorem{lemma}{\indent Lemma}
\newtheorem{theorem}{\indent \emph{\textbf{Theorem}}}
\newtheorem{corollary}{\indent \textbf{\emph{Corollary}}}

\begin{document}

\makeatletter
\newcommand{\ud}{\mathrm{d}}
\newcommand{\rmnum}[1]{\romannumeral #1}
\newcommand{\Rmnum}[1]{\expandafter\@slowromancap\romannumeral #1@}
\makeatother

\preprint{APS/123-QED}

\title{Asymptotic Quantum Algorithm for the Toeplitz Systems}

\author{ Lin-Chun Wan}
\author{ Chao-Hua Yu}
\author{ Shi-Jie Pan}
\affiliation{State Key Laboratory of Networking and Switching Technology, Beijing University of Posts and Telecommunications, Beijing, 100876, China}
\author{Fei Gao}
\email{gaof@bupt.edu.cn}
\affiliation{State Key Laboratory of Networking and Switching Technology, Beijing University of Posts and Telecommunications, Beijing, 100876, China}
\author{Qiao-Yan Wen}
\author{Su-Juan Qin}
\affiliation{State Key Laboratory of Networking and Switching Technology, Beijing University of Posts and Telecommunications, Beijing, 100876, China}

\date{\today}

\begin{abstract}
Solving the Toeplitz systems, which is to find the vector $x$ such that $T_nx = b$ given an $n\times n$ Toeplitz matrix $T_n$ and a vector $b$, has a variety of applications in mathematics and engineering. In this paper, we present a quantum algorithm for solving the linear equations of Toeplitz matrices, in which the Toeplitz matrices are generated by discretizing a continuous function. It is shown that our algorithm's complexity is nearly  $O(\kappa\textrm{log}^2 n)$, where $\kappa$ and $n$ are the condition number and the dimension of $T_n$ respectively. This implies our algorithm is exponentially faster than the best classical algorithm for the same problem if  $\kappa=O(\textrm{poly}(\textrm{log}\,n))$. Since no assumption on the sparseness of $T_n$ is demanded in our algorithm, it can serve as an example of quantum algorithms for solving non-sparse linear systems.
\begin{description}
\item[PACS numbers]
03.67.Dd, 03.67.Hk
\end{description}
\end{abstract}

\pacs{Valid PACS appear here}
\maketitle


\section{Introduction}

Quantum information processing has been shown enormously advantageous in preserving security and privacy in communication and information retrieval \cite{bennett2014quantum,wei2017generic}, as well as in computing for solving certain problems \cite{alg,yu2016quantum}. In recent years, since the first example --- quantum algorithm for linear systems of equations (HHL algorithm) --- was presented by Harrow et al. \cite{eq1}, a number of quantum algorithms for other problems in numerical computation, such as linear regression \cite{fit1,yu2017quantum}, poisson equation \cite{pos}, and finite element method \cite{finite}, have been subsequently proposed with significant speedup over their classical counterparts. These works motivate us to design fast quantum algorithms for solving more problems in numerical computation.

One of the most important problems in numerical computation is solving linear equations with a Toeplitz matrix. Solving this linear systems has a variety of applications in many areas of science and engineering, such as signal processing \cite{sign}, time series analysis \cite{time}, image restoration problems \cite{image}, queueing problems \cite{queue}, minimum realization problems in control theory \cite{application} and numerical integration problems \cite{integral}. In general, the Toeplitz systems are obtained by discretization of continuous problems and the dimension $n$ is related to the grid parameter of the discretization. More specifically, the given Toeplitz matrices $T_n$ are generated by a generating function $f$, i.e., the elements of every diagonal of $T_n$ are given by the Fourier coefficients of $f$. Therefore, the linear systems usually are of very large dimensions so that more efficient algorithms for solving these systems deserve to be explored \cite{ill1}.

In the past decades, many scholars have paid their attention to developing the methods for speeding up solving Toeplitz systems. A number of advanced methods have been presented, such as fast direct methods \cite{Superfast}, iterative methods \cite{CGT}, and circulant approximation methods \cite{cirapp}. There is a wonderful treasury of classical algorithms for solving Toeplitz systems. Nevertheless, since the time complexity of these methods are $\Omega(n)$, it is still a hard work to tackle Toeplitz systems with very large $n$ on a classical computer.

As of now, some work regarding Toeplitz matrices has been done in the quantum setting. In 2016, A. Mahasinghe and J.B.Wang presented an efficient quantum algorithm for implementing sparse or Fourier-sparse Toeplitz matrices \cite{Tb}. Whereafter, the algorithm presented in \cite{zhou2017efficient} provides a better way to implement Toeplitz matrices, requiring few resources and without the sparsity assumption on $T_n$. These two algorithms have some significant applications in physics, mathematics and engineering related field. It should be noted that, although \cite{zhou2017efficient} has shown how to implement $T_n|b\rangle$ by embedding $T_n$ in a larger circulant matrix (a special kind of Toeplitz matrices), as well as how to invert a circulant matrix efficiently, it is easy to see inverting $T_n$ cannot be realized by this trick. However, as mentioned above, many problems can be transformed into solving Toeplitz systems. That is to say, in these practical applications, what is really relevant is the inverse of such matrices rather than the matrices themselves. This is exactly what we focus on in this paper.

Specifically, we present a quantum algorithm to solve the Toeplitz systems, i.e., finding a quantum state $|x\rangle=\sum_ix_i|i\rangle/\|\sum_ix_i|i\rangle\|$ satisfying the linear equations, $T_nx = b$, in which the vector $b$ is given by $|b\rangle=\sum_ib_i|i\rangle/\|\sum_ib_i|i\rangle\|$, $T_n$ is generated by a continuous function and not limited to sparse. The basic idea of our algorithm is that, due to the fact that the Toeplitz matrices can be well approximated by some easier-to-tackle circulant matrices, our quantum algorithm solves the linear equations of Toeplitz matrices by resorting to solve that of the circulant matrices. In addition, we make full use of the relationship between circulant matrices and the generating function, so that we can directly acquire the eigenvalues of the circulant matrices by computing corresponding value of the generating function. Thus it can avoid performing the phase estimation which requires higher complexity to reveal corresponding eigenvalues. It is shown that our algorithm is exponentially faster than the classical methods when the generating functions $f$ are strictly positive continuous real-valued functions and the Toeplitz matrices $T_n$ are well-conditioned. We call Toeplitz matrices well-conditioned when $\kappa= O(\textrm{poly}(\textrm{log}\,n))$. Moreover, since the solution is encoded  in the final quantum state of our algorithm, it is advantageous to extract some interesting features of the solution and it can also be used as an ingredient in other quantum algorithms.

The rest of this paper is organized as follows. In Sec.II, we review some basic concepts and properties of Toeplitz matrices. And we describe the details of our quantum algorithm in Sec.III. The error and runtime analysis of this algorithm are given in Sec.IV. We then discuss some special cases in Sec.V. And the last section draws the conclusions.


\section{A review about Toeplitz matrices}

A Toeplitz matrix $T_n$ is a matrix of size $n\times n$ whose coefficients along each diagonal are constant. More precisely, a Toeplitz matrix has the form
\begin{equation}\label{eq:defineT}
T_n =
\left( \begin{array}{ccccc}
t_{0} & t_{-1} & t_{-2} & \ldots & t_{-(n-1)}\\
t_{1} & t_{0} & t_{-1} & \ddots & \vdots \\
t_{2} & t_{1} & t_{0} & \ddots & t_{-2}  \\
\vdots & \ddots & \ddots & \ddots & t_{-1}\\
\ t_{(n-1)} & \ldots & t_{2} & t_{1} & t_{0}\\
\end{array} \right)
\end{equation}
where $t_{k,j} = t_{k-j}$, thus, it can be determined by a sequence $\{t_k\}_{k=-n+1}^{n-1}$ with only $2n-1$ entries.

As we mentioned above, in many applications, the Toeplitz matrices are obtained by discretization of continuous problems. More explicitly, let $C_{2\pi}^+$ be the set of all $2\pi$-periodic strictly positive continuous real-valued functions defined on $[0,2\pi]$. For all $f\in C_{2\pi}^+$,
\begin{equation}
t_k=\frac{1}{2\pi}\int_{0}^{2\pi}\!\!\!f(\lambda)e^{-ik\lambda}\ud\lambda, \qquad k=0,\pm1,\pm2,\cdots.
\end{equation}
are the Fourier coefficients of $f$. Then, let $T_n(1\leq n<\infty)$ be the sequence of Toeplitz matrices whose entries along the $k$-th diagonal are $t_k$. The function $f$ is called the generating function of the sequence of Toeplitz matrices $T_n$, see \cite{f}, and the sequence of matrices $T_n$ is often denoted as $T_n(f)$. We are interested in solving the Toeplitz systems $T_n(f)x=b$.

The properties of Toeplitz matrices are well known and easily derived. We describe a simple version here, for details, see \cite{Toe}. The Toeplitz matrices $T_n(f)$ will be Hermitian if $f$ is real-valued function. In fact,
\begin{equation*}
\begin{aligned}
t^*_k&=\frac{1}{2\pi}\int_{0}^{2\pi}f^*(\lambda)e^{ik\lambda}\ud\lambda\\
&=\frac{1}{2\pi}\int_{0}^{2\pi}f(\lambda)e^{ik\lambda}\ud\lambda\\
&=t_{-k}.
\end{aligned}
\end{equation*}

Another useful property is that  when $T_n(f)$ is Hermitian, let $\lambda_k$ be the eigenvalues of a Toeplitz matrix $T_n(f)$, then
\begin{equation*}
f_{min}\leq\lambda_k\leq f_{max},
\end{equation*}
where $f_{min},f_{max}$ represent the smallest value and the largest value of $f$ respectively.
In particular,
\begin{equation*}
\begin{aligned}
&\lim_{n\rightarrow\infty}\max_{k}\lambda_k=f_{max},\\
&\lim_{n\rightarrow\infty}\min_{k}\lambda_k=f_{min}.
\end{aligned}
\end{equation*}
Note that the generating function $f$ is strictly positive then the Toeplitz matrix $T_n(f)$ is nonsingular. Another thing worth emphasizing is that in practical applications, we are often given the generating functions $f$ instead of the Toeplitz matrices $T_n(f)$. Typical examples of generating functions are \cite{sign,image,CGT,f,f1}.

There is a common special case of Toeplitz matrix when every row of the matrix is a right cyclic shift of the row above it. In this case, the structure becomes
\begin{equation}
C_n =
\left( \begin{array}{ccccc}
c_{0}       & c_{1}     & c_{2}      & \ldots      & c_{(n-1)}\\
c_{(n-1)}   & c_{0}     & c_{1}      & \ldots    & c_{(n-2)}\\
c_{(n-2)}   &           & \ddots     & \ddots      &  \vdots  \\
\vdots      &           & \ddots     &             & c_{1}\\
\ c_{1}     & \ldots    &            & c_{(n-1)}   & c_{0}\\
\end{array} \right).
\end{equation}
A matrix of this form is called a $circulant$ matrix. The following theorem summarizes the properties regarding eigenvalues and eigenvectors of circulant matrices, and more details can be found in \cite{Toe}.

\begin{theorem}[\cite{Toe}] \label{theorem:1}
Every circulant matrix $C_n$ can be diagonalized by the Fourier matrix $F_n$. That means it has the form $C_n=F_n^\dagger\Lambda_n F_n$, where the entries of $F_n$ are given by
$$ [F_n]_{j,k}=\frac{1}{\sqrt{n}}e^{-2\pi ijk/n},\quad 0\leq j,k\leq n-1,$$
and $\Lambda_n$ is diagonal matrix corresponding eigenvalues are given by
\begin{equation}\label{eq:definelambda}
\psi_m=\sum_{k=0}^{n-1}c_ke^{-2\pi imk/n} \qquad m=0,1,\dots,n-1,
\end{equation}
\end{theorem}
Apparently, $C_n$ is normal.

\begin{corollary}[\cite{Toe}] \label{corollary:1}
Let $C$ and $B$ be $n\times n$ circulant matrices with eigenvalues
$$
\psi_m=\sum_{k=0}^{n-1}c_ke^{-2\pi imk/n},\beta_m=\sum_{k=0}^{n-1}b_ke^{-2\pi imk/n}
$$
respectively. Then\\
(1) $C$ and $B$ commute and
\begin{equation}
CB=BC=F_n^\dagger\gamma F_n,
\end{equation}
where $\gamma = \mathrm{diag}(\psi_m\beta_m),$ and $CB$ is also a circulant  matrix.\\
(2) $C+B$ is a circulant matrix and
\begin{equation}\label{eq:csum}
C+B=F_n^\dagger\Omega F_n,
\end{equation}
where $\Omega=\mathrm{diag}\{(\psi_m+\beta_m)\}$.\\
(3) If $\psi_m\not=0; m=0,1,\dots,n-1,$ then $C$ is nonsingular and
\begin{equation}\label{eq:multiply}
C^{-1}=F_n^\dagger\Lambda_n^{-1}F_n.
\end{equation}
\end{corollary}

One technique to solve the problem involving Toeplitz matrices is to construct a sequence of circulant matrices which are asymptotically equivalent to the Toeplitz matrices. Obviously the choice of constructing a sequence of circulant matrices to approximate the sequence of Toeplitz matrices is not unique, therefore we need to choose a construction of which properties are most desirable. It will prove useful to adopt the circulant matrices defined in \cite{f,Toe}. In particular, define $C_n(f)$ to be the circulant matrix with top row $(c_0,c_1 ,\cdots, c_{n-1})$ where
\begin{equation}\label{eq:definec}
c_k=\frac{1}{n}\sum_{j=0}^{n-1}f(2\pi j/n)e^{2\pi ijk/n}.
\end{equation}
According to Theorem \ref{theorem:1}, the eigenvalues of $C_n(f)$ are simply $f(2\pi m/n)$:
\begin{equation}\label{eq:defineclambda}
\begin{aligned} 
\psi_{m}&=\sum_{k=0}^{n-1}c_k e^{-2\pi imk/n}\\
          &=\sum_{k=0}^{n-1}\Big(\frac{1}{n} \sum_{j=0}^{n-1}f(2\pi j/n)e^{2\pi ijk}\Big)e^{-2\pi imk}  \\
          &=\sum_{j=0}^{n-1}f(2\pi j/n)\Bigg\{ \frac{1}{n} \sum_{k=0}^{n-1}e^{2\pi i(j-m)k} \Bigg\}\\
          &=f(2\pi m/n) \qquad   m=0,1,\ldots,n-1
\end{aligned}
\end{equation}
by using the orthogonality of the complex exponentials \cite{Toe}. The $C_n(f)$ is called associated circulant matrices of $T_n(f)$ \cite{square}. To put it simple, the associated circulant matrices have the form
\begin{equation*}
C_n(f)=F_n^\dagger\Lambda_n F_n
\end{equation*}
where
\begin{equation*}
\Lambda_n=\textrm{diag}\{f(0),f(2\pi/n),\ldots,f(2\pi (n-1)/n)\}.
\end{equation*}

It has been observed that in many applications substituting $T_n(f)$ with $C_n(f)$ often leads to very useful and dramatic simplification \cite{square}. Our quantum algorithm will also follow this idea to achieve exponential speedup over the classical algorithm.

\section{Quantum Algorithm}

Now, we design the quantum algorithm to solve the linear system of associated circulant matrices. In particular, from (\ref{eq:multiply}) and (\ref{eq:defineclambda}), we know $C_n^{-1}(f)=F_n^\dagger\Lambda_n^{-1}F_n$ where $\Lambda_n^{-1}$ is a diagonal matrix with corresponding eigenvalue $1/f(2\pi j/n)$. Thus, the essential problem is how to implement $\Lambda_n^{-1}$. It is a natural candidate for applying the HHL algorithm \cite{eq1}. Surprisingly, because of the distinctive structure of the associated circulant matrices, we can bypass the phase estimation.

As argued in most of quantum algorithms, we always assume that the implementation of initial state $|b\rangle$ is efficient, since our algorithm could be a subroutine in a larger quantum algorithm of which some other component is responsible for producing $|b\rangle$. In general, preparing an arbitrary initial quantum state is really challenging. Only a few implementation schemes have been presented \cite{prepare,prepare1}. Besides, one can use the quantum Random Access Memory (qRAM) \cite{prepare2} to provide $|b\rangle$. We are looking forward to more outstanding work in this field.

To extract the eigenvalues, we assume there is an oracle that accesses the values of generating function $f$. Specifically it allows us to perform the map
$$\sum_{j=0}^{n-1}b_j|j \rangle \xrightarrow{oracle}\sum_{j=0}^{n-1}b_j|j \rangle|f(2\pi j/n)\rangle.$$
We don't consider the computation complexity of the oracle since the generating function $f$ is always efficiently computable. In fact, it can be implemented efficiently by using the quantum circuit model. The section 3.2.5 of \cite{oracle1} show that given a classical circuit for computing $f$ there is a quantum circuit of comparable efficiency  which computes the transformation $U_f$ on a quantum computer. And some quantum algorithms and circuits of fundamental numerical functions proposed by Bhaskar et. al. \cite{oracle} may be helpful to implement the $U_f$. For our purposes, we can regard it as a black box.

The specific process of our quantum algorithm is summarized by the following five steps.\\
\textbf{Algorithm:}\\
1. Apply Fourier transform on $|b\rangle$ and denote the outputed state by $|b'\rangle$.\\
2. Decompose $|b'\rangle$ in the computational basis, i.e., $|b'\rangle=\sum_{j=0}^{n-1}b_j|j \rangle$,
and use the oracle to obtain
$$\sum_{j=0}^{n-1}b_j|j \rangle|f(2\pi j/n)\rangle.$$
The query complexity to prepare this state is $O(1)$.\\
3. Add a qubit and perform a controlled-rotation on $|f(2\pi j/n)\rangle $ yields
$$\sum_{j=0}^{n-1}b_j|j \rangle|f(2\pi j/n)\rangle\Bigg(\sqrt{1-\frac{m^2}{f^2(2\pi j/n)}}|0\rangle+\frac{m}{f(2\pi j/n)}|1\rangle\Bigg)$$
where $m$ is an appropriate constant that $m\leq\textrm{min}_j|\psi_j|$, and $\psi_j$ are the eigenvalues of $C_n(f)$.\\
4. Uncompute the second qubit and use the amplitude amplification \cite{amplitude} on the last register to obtain $|1\rangle$, hence we will have the following state with higher probability
$$\sqrt{\frac{1}{\sum_{j=0}^{n-1}m^2|b_j|^2/|f(2\pi j/n)|^2}}\sum_{j=0}^{n-1}b_j\frac{m}{f(2\pi j/n)}|j\rangle$$
which is proportional to $\Lambda^{-1}|b'\rangle=\sum_{j=0}^{n-1}b_j/f(2\pi j/n)|j\rangle$ up to normalization, and denoted by $|b^*\rangle$.\\
5. Perform the inverse Fourier transform and get
$$ |x^*\rangle=F_n^\dagger|b^*\rangle.$$

The $|x^*\rangle$ is the quantum state that we desired, and in the following section we will prove it is close to the normalized solution $|x\rangle=\frac{T_n^{-1}(f)|b\rangle}{\|T_n^{-1}(f)|b\rangle\|}$ of the Toeplitz system.

\section{Error Analysis and Runtime}

In this section we will show that the error of final state of our algorithm to the ideal state can always be small enough. Furthermore, we demonstrate two significative corollaries to ascertain the magnitude of error $\epsilon$ for some certain cases. And we also perform analysis on the time complexity of our quantum algorithm and show its runtime advantage over the classical algorithm.

As our main idea is substituting $T_n(f)$ with $C_n(f)$, in order to make such a substitution meaningful, the sequence of Toeplitz matrices need to converge to their associated circulant matrices. We define the matrices convergence as the following theorem.

\begin{theorem}\label{theorem:4}
Let $T_n(f)$ be a sequence of Toeplitz matrices generated by a strictly positive continuous real-valued function, $C_n(f)$ is the associated circulant matrices defined in (\ref{eq:definec}), then
\begin{equation}\label{eq:th4}
\lim_{n\rightarrow\infty}\frac{\|T_n(f)-C_n(f)\|_F}{\|T_n(f)\|_F}=0
\end{equation}
where$\|A\|_F$ denotes the Frobenius norm of a matrix $A$.\\
\end{theorem}
Proof: see the appendix A.

Let
\begin{equation*}
x^*=C_n^{-1}(f)b\qquad  x=T_n^{-1}(f)b
\end{equation*}
\begin{equation*}
|x^*\rangle=\frac{C_n^{-1}(f)|b\rangle}{\|C_n^{-1}(f)|b\rangle\|}\qquad
|x\rangle=\frac{T_n^{-1}(f)|b\rangle}{\|T_n^{-1}(f)|b\rangle\|}
\end{equation*}
where $\|x\|=\sqrt{x^{H}x}$ denotes the 2-norm of a vector $x$.

Recalling theorem \ref{theorem:4}, for a given $\epsilon$, we can always choose an $n$ large enough, so that
\begin{equation*}
\frac{\|T_n(f)-C_n(f)\|_F}{\|T_n(f)\|_F}\leq \epsilon
\end{equation*}
then, from relevant conclusion of \cite[Sec. 5.8]{maxcpt}, we know that
\begin{equation}\label{eq:errork}
\frac{\|x^*-x\|}{\|x\|}=\frac{\|C_n^{-1}(f)b-T_n^{-1}(f)b\|}{\|T_n^{-1}(f)b\|}\leq\frac{\epsilon\kappa}{1-\epsilon\kappa}
\end{equation}
where $\kappa$ is the condition number of $T_n(f)$, and
\begin{equation}\label{eq:ferror}
\begin{aligned}
&\|\,|x^*\rangle-|x\rangle\|\\
&=\Big\|\frac{C^{-1}|b\rangle}{\|C^{-1}|b\rangle\|}-\frac{T^{-1}|b\rangle}{\|T^{-1}|b\rangle\|}\Big\|\\
&=\Big\|\frac{(\|T^{-1}|b\rangle\|-\|C^{-1}|b\rangle\|)C^{-1}|b\rangle}{\|C^{-1}|b\rangle\|\,\|T^{-1}|b\rangle\|}-\frac{T^{-1}|b\rangle-C^{-1}|b\rangle}{\|T^{-1}|b\rangle\|}\Big\|\\
&\leq\frac{\big|\,\|T^{-1}|b\rangle\|-\|C^{-1}|b\rangle\|\,\big|}{\|T^{-1}|b\rangle\|}+\frac{\|T^{-1}|b\rangle-C^{-1}|b\rangle\|}{\|T^{-1}|b\rangle\|}\\
&\leq2\frac{\|T^{-1}|b\rangle-C^{-1}|b\rangle\|}{\|T^{-1}|b\rangle\|}\\
&=2\frac{\|T^{-1}b-C^{-1}b\|}{\|T^{-1}b\|}
\leq\frac{2\epsilon\kappa}{1-\epsilon\kappa}
\end{aligned}
\end{equation}
for simplicity, we write $C_n^{-1}(f)$ and $T_n^{-1}(f)$ as $C^{-1}$ and $T^{-1}$ respectively in this inequality.
Apparently, the output of our algorithm can approximate the normalized solution of a well-conditioned Toeplitz system with desired precision as long as we choose an $n$ large enough.

The above process proves the correctness of the algorithm in a certain sense. Naturally, we need to seek ways to provide error bound estimates. Depending on different properties of associated sequence $\{t_k\}$, we get different error estimates as shown in the following corollaries.

\begin{corollary}\label{corollary:2}
Let $T_n(f)$ be a sequence of well-conditioned Toeplitz matrices generated by a strictly positive continuous real-valued function. If the associated series $\sum_{k=0}^{\infty}|t_k|$ is a convergent P-series, then we can acquire a normalized solution of corresponding Toeplitz system with error $\epsilon=O(\frac{\textrm{ln}n\textrm{log}n}{n})$.
\end{corollary}
Proof: see the appendix B.

It is well known that the magnitude of Fourier coefficients of most of primitive functions is $o(\frac{1}{n})$, where $o$ denotes the higher order infinitely small quantity, i.e., their associated series $\sum_{k=0}^{\infty}|t_k|$ are convergent P-series. Therefore, the error of a lot of Toeplitz systems can be estimated conveniently from this corollary. Moreover, some useful judgment theorems about the magnitude of Fourier coefficients of functions are listed in \cite{hardy1999fourier}, when the objective generating functions satisfy the corresponding conditions, the final error will also be bounded by $O(\frac{\textrm{ln}n\textrm{log}n}{n})$.

\begin{corollary}\label{corollary:3}
Let $T_n(f)$ be a sequence of Toeplitz matrices generated by a strictly positive continuous real-valued function. If the associated sequence $\{t_k\}$ sastisfy $\sum^{\infty}_{k=-\infty}|kt_k|<\infty$ and the spectrum norm $\|T_n\|\leq 1$, for a vector with the form
\begin{equation*}
  b=(0,\ldots,0,b_{-L},\ldots,b_0,\ldots,b_L,0,\ldots,0),
\end{equation*}
we can acquire a normalized solution of corresponding Toeplitz system with error $\epsilon=O(1/\sqrt{n})$.
\end{corollary}
Proof: see the appendix C.

This corollary is based on the results of literature \cite{square}. It extend the convergence theorem to a form of which $b$ having only a finite number of nonzero term. And a large class of communication receiver design problems involving a similar form can be solved efficiently.

Based on these corollaries, we can conveniently estimate the scale of $\epsilon$ for a specific Toeplitz system. This seems to be different from the normal way we design and analyze algorithms of which the accuracy $\epsilon$ and dimension $n$ are given beforehand. We would like to emphasize that the Toeplitz systems are often given by a continuous function. And we are required to design an algorithm to solving the problem with accuracy $\epsilon$. In the actual operation, we first discretize the continuous function and then solve this discrete system. Thus, the dimension $n$ is not fixed but operational. And it depends on the desired accuracy and can be determined by the above corollaries.

Reviewing our algorithm, the (inverse) quantum Fourier transform takes time $O(\textrm{log}^2n)$, and
the cost of invoking the oracle is $O(1)$ while the computation complexity of the oracle can be ignored.
Then, we consider the success probability of the post-selection in the process of implementing $\Lambda_n^{-1}$.
Since $m\leq\textrm{min}_j|\psi_j|$, for choosing $m=f_{min}$, the success probability is $\Omega (1/\mu^2)$, where $\mu=f_{max}/f_{min}$, and $O(\mu^2)$ measurements is required. Using amplitude amplification, we need only repeat $O(\mu)$ times. Putting these all together, our quantum algorithm takes time $O(\mu\textrm{log}^2n)$. Noting that the condition number $\kappa$ approximate $f_{max}/f_{min}$ when $n$ getting large, the time complexity of our algorithm is also nearly $O(\kappa \textrm{log}^2n)$.

Replacing Toeplitz matrices with their associated circulant matrices has been widely used to solve the problems involving the inverses of Toelplitz matrices. It takes two steps to complete this process by classical algorithms: (i) compute the top row of the circulant matrix. (ii) solve the linear system of circulant matrices. Both of these steps can be performed efficiently. In particular, the top row can be computed in $O(n)$ \cite{Toe}. And according to Theorem \ref{theorem:1}, the linear systems of circulant matrices can be solved in $O(n\textrm{log}n)$ using the fast Fourier transform. Therefore, our quantum algorithm is exponentially faster than the corresponding classical algorithm when the condition numbers of the Toeplitz matrices are $O(\textrm{poly}(\textrm{log}\,n))$.

\section{Special case}

Sometimes, there might be some problems in practical situations that we only know the sequence of Toeplitz matrices $T_n$ but not the generating function $f$. Fortunately, it will often begin with Toeplitz matrices in the Wiener class \cite{Toe}. A sequence of Toeplitz matrices for which the $\{t_k\}$ are absolutely summable is said to be in the Wiener class, i.e.,  the infinite sequence $\{t_k;k=\dots,-2,-1,0,1,2,\dots\}$ which defines the matrices $T_n$  meet
\begin{equation}\label{eq:abssum}
\sum_{k=-\infty}^{\infty}|t_k|<\infty.
\end{equation}

As mentioned in \cite{Toe}, one natural idea for estimating the eigenvalues is to approximate the generating function by:
\begin{equation}\label{eq:newf1}
\hat{f}_n(\lambda)=\sum_{k=-(n-1)}^{n-1}t_ke^{ik\lambda}, \qquad \lambda\in[0,2\pi]
\end{equation}

On the one hand, the function $\hat{f}_n(\lambda)$ define a circulant matrices sequence $C_n(\hat{f}_n)$ by (\ref{eq:definec}). Note that since the $\{t_k\}$ are absolutely summable, they are also square summable
\begin{equation*}
\sum_{k=-\infty}^{\infty}|t_k|^2\leq\Bigg\{\sum_{k=-\infty}^{\infty}|t_k|\Bigg\}^2<\infty.
\end{equation*}
Thus, according to the proof in appendix A, it can be seen that the circulant matrices $C_n(\hat{f}_n)$ converges to $T_n(f)$ in the form (\ref{eq:th4}). Based on this convergence theory, our quantum algorithm is still feasible.

On the other hand, the eigenvalues of circulant matrices $C_n(\hat{f}_n)$ are
\begin{equation*}
\begin{aligned}
\hat{f}_{n}(2\pi j/n)&=\!\!\!\!\!\sum_{k=-(n-1)}^{n-1}\!\!\!\!\!t_ke^{2\pi ijk/n}&\\
                     &=\sum_{k=0}^{n-1}t_ke^{2\pi ijk/n}+\sum_{k=0}^{n-1}t_{-k}e^{-2\pi ijk/n}-t_0\\
\end{aligned}
\end{equation*}
where $j=0,1,\ldots,n-1.$ Apparently, they can be seen as the results of performing discrete Fourier transformation on the sequences $\{t_k\}$ and then subtracting a extra $t_0$.

In order to solve these Toepltiz systems, we first rescale the matrices $T_n$ by the factor $f_{max}$, such that $\lambda_{k}\in[1/\mu,1]$. Then we call the algorithm presented in \cite{Zhou2016Quantum} to perform Fourier transform and encode the Fourier coefficients in the computational basis. More formally, the quantum Fourier transform in the computational basis (QFTC) can perform the transformation:
\begin{equation*}
  |j\rangle\xrightarrow{QFTC}|j\rangle|y_j\rangle
\end{equation*}
where $y_j=\sum_{k=0}^{n-1}t_ke^{2\pi ijk/n}$ in this context.

Since the $|j\rangle$ is only used to control the application of quantum operators acting on other registers, we can do the similar operations in the additional registers and get
\begin{equation*}
  \sum_{j=0}^{N-1}b_j|j\rangle|0\rangle|0\rangle|t_0\rangle\rightarrow\sum_{j=0}^{N-1}b_j|j\rangle|y_j\rangle|y'_j\rangle|t_0\rangle
\end{equation*}
where $y'_j=\sum_{k=0}^{n-1}t_{-k}e^{-2\pi ijk/n}$.
Then, calculate $y_j+y'_j-t_0$ by the quantum adder \cite{draper2000addition} and encode the result in another registers:
\begin{equation*}
  \sum_{j=0}^{N-1}b_j|j\rangle|y_j\rangle|y'_j\rangle|t_0\rangle|\hat{f}_{n}(2\pi j/n)\rangle
\end{equation*}
Finally, uncompute the ancillas to obtain
\begin{equation*}
  \sum_{j=0}^{n-1}b_j|j\rangle|Anc_j\rangle|\hat{f}_{n}(2\pi j/n)\rangle\rightarrow \sum_{j=0}^{n-1}b_j|j\rangle|\hat{f}_{n}(2\pi j/n)\rangle
\end{equation*}

The above process actually completes the step 2 of our algorithm, i.e.,
\begin{equation*}
  \sum_{j=0}^{n-1}b_j|j\rangle\xrightarrow{oracle} \sum_{j=0}^{n-1}b_j|j\rangle|\hat{f}_{n}(2\pi j/n)\rangle.
\end{equation*}
Proceeding to the next steps of the algorithm, where $m=O(1/\mu)$, we can get a  quantum state approximating the solution of the Toeplitz system.

It is worth noting that the algorithm QFTC requires an oracle $O_t$, where $O_t|0\rangle=\sum_{k=0}^{n-1}t_k|k\rangle$. The oracle can be efficiently implemented if $\{t_k\}_{k=0}^{n-1}$ is efficiently computable or using the qRAM \cite{prepare2}. And the QFTC can be performed to accuracy $\epsilon$ with fidelity $1-\delta$ using $O(\textrm{log}^2n/(\delta\epsilon))$ one- or two-qubit gates. According to the error analysis in \cite{eq1}, taking the error as $O(\epsilon_0/\mu)$ in computing $\hat{f}_{n}(2\pi j/n)$ induces a final error $\epsilon_0$ of $|x^*\rangle$. Thus, employing the QFTC as a subroutine to extract the eigenvalues, the complexity of our quantum algorithm is nearly $O[\kappa^2\textrm{log}^2n/(\delta\epsilon_0)]$.  This suggests that our algorithm is exponentially fast when $1/\epsilon_0,\kappa = O(\textrm{poly}(\textrm{log}\,n))$.

\section{Discussion}

We notice that some meaningful results about the asymptotic equivalence of Toeplitz matrices and the associated circulant matrices have been presented recently \cite{zhu2017asymptotic}. These results establish the individual asymptotic convergence of the eigenvalues between $T_n$ and $C_n$ while the theorem \ref{theorem:4} of this paper actually characterizes the certain collective asymptotic behaviors of the eigenvalues. The individual asymptotic equivalence seems to be stronger than collective asymptotic equivalence. We describe these results here to demonstrate that our algorithm might achieve faster convergence rate if the certain conditions on $\{t_k\}$ are met.
\begin{theorem}[\cite{zhu2017asymptotic}]\label{Convergence of eigenvalues}
Suppose that the  sequence $\{t_k\}$ is absolutely summable, $C_n$ is the associated circulant matrices of the Toeplitz matrices $T_n$. Then
\begin{equation}
\lim_{n\rightarrow\infty} \max_{l\in n}|\lambda_l(T_n)-\lambda_{\rho(l)}(C_n)|=0
\end{equation}
where $\lambda_l(T_n)$ are the eigenvalues of $T_n$ satisfying $\lambda_0(T_n)\geq\cdots\geq\lambda_{n-1}(T_n)$, and
$\lambda_{\rho(l)}(C_n)$ are the eigenvalues of $C_n$ satisfying $\lambda_{\rho(0)}(C_n)\geq\cdots\geq\lambda_{\rho(n-1)}(C_n)$.
\end{theorem}
\begin{theorem}[\cite{zhu2017asymptotic}]\label{Convergence of eigenvalues of band Toeplitz}
Suppose that the $t_k=0$  for all $|k|>r$, i.e., $T_n$ is a band Toeplitz matrix when $n>r$. Then
\begin{equation}
\max_{l\in n}|\lambda_l(T_n)-\lambda_{\rho(l)}(C_n)|=O(\frac{1}{n})
\end{equation}
as $n\rightarrow\infty$.
\end{theorem}

In addition, because the kernel idea of our algorithm is adopting associated circulant matrices to substitute the Toeplitz matrices, it must be intractable to solve the ill-conditioned Toeplitz system. Evidently, the error of the solution may not be controlled from (\ref{eq:errork}) when the condition number of $T_n$ is unbounded. Fortunately, in some cases, when the $|b\rangle$ is in the well-conditioned part of $T_n$ (i.e., the subspace spanned by the eigenvectors with large eigenvalues), we can also implement the invertion. The corresponding process have been demonstrated in detail in \cite{eq1}. Another way to handle ill-conditioned Toeplitz systems is to precondition the Toeplitz matrices. A number of preconditioners developed for ill-conditioned Toeplitz systems have been presented, including Band-Toeplitz Preconditioners \cite{band} and Circulant Preconditioners \cite{ill2}. Since the product of two circulant matrices is a circulant matrix, and the product of circulant matrix times a vector is available by our quantum algorithm with a little change. Our algorithm can run much faster with a suitable circulant preconditioner even for ill-conditioned Toeplitz systems. Moreover, it is an interesting problem that how to establish the quantum version of these preconditioners. And that is our next work.

\section{Conclusion}

Solving the Toeplitz systems plays a pivotal role in many areas of science and engineering. We have addressed this problem in the quantum settings and proposed an efficient quantum algorithm to solve it. Taking advantages of quantum computation and the structure of Toeplitz matrices, our algorithm achieves exponential speedup over classical algorithms for the well-conditioned Toeplitz matrices.

Besides solving linear equations with a Toeplitz matrix, our quantum algorithm can deal with many other problems regarding the approximation of a Toeplitz matrix by its associated circulant matrix. In fact, many different convergence forms, such as weak convergence form \cite{Toe}, finite-term quadratic form\cite{square}, of approximation between Toeplitz matrices and their associated circulant have been exploited for different applications. Although the matrix convergence forms are different, similar to our  algorithm, all of these methods adopt the idea that substituting the Toeplitz matrices with their associated circulant matrices to solve corresponding problems. Therefore, the common methodology makes it be possible that our quantum algorithm can be successfully applied to these applications.

Solving linear systems of equations is a fundamental problem that arises frequently in science and engineering. The quantum algorithms for solving sparse linear systems have been well studied \cite{eq1,childs2015quantum,clader2013preconditioned}. More recently, L. Wossnig, Z. Zhao, A. Prakash presented a quantum linear system algorithm for dense matrices based on a new data structure which prepares quantum states corresponding to the rows and the vector of Euclidean norms of the rows of the matrices \cite{wossnig2017quantum}. The algorithm achieved a polynomial improvement over known quantum linear system algorithms when the dense matrix with spectral norm bounded by a constant. Simply applying this algorithm to solve the Toeplitz systems gives only a polynomial improvement. Thus, designing efficient quantum algorithms for solving non-sparse linear systems with  certain special structures remains a crucial challenge. Our work is a significant addition for this direction. And it is an interesting open question if one can achieve a exponential improvement for solving Toeplitz systems in the model of \cite{wossnig2017quantum} or given a black-box access to the matrix elements.

What's more, Ke Ye and Lek-Heng Lim pointed out every matrix is a product of Toeplitz matrices \cite{product}, their conclusion that every $n \times n$ matrix can be decomposed into $\lfloor n/2\rfloor+1$ Toeplitz matrices is exciting. That is if a large non-sparse matrix $A$ has a known Toeplitz decomposition, one can solve the corresponding linear systems with lower time complexity by a high-efficiency quantum algorithm for Toeplitz systems (though we still do not know how to compute Toeplitz decompositions efficiently). It should be noted that our algorithm is not suitable for this method because of the asymptotic equivalence, but this method opens up a new horizon in solving general linear equations and deserves further investigation.

\appendix
\section{Proof of the Theorem \ref{theorem:4}}

In this appendix, we prove the conclusion in Theorem \ref{theorem:4}. For simplicity, we write $\|A\|_F$ as $|A|$ in the following processes. And we firstly give some involved theories.

\begin{lemma}\label{definition:1}
The Frobenius norm of an $n\times n$ matrix $A=[a_{k,j}]$ is equal to
\begin{equation}\label{eq:Fnorm}
\begin{aligned}
\|A\|_F&=\Bigg(\sum_{k=0}^{n-1}\sum_{j=0}^{n-1}|a_{k,j}|^2\Bigg)^{1/2}\\
       &=[tr(A^*A)]^{1/2}=\Bigg[\sum_{k=0}^{n-1}\lambda_k(A^*A)\Bigg]^{1/2}
\end{aligned}
\end{equation}
\end{lemma}

\begin{lemma}\label{theorem:3}
Let $A$ be a matrix with eigenvalues $ \lambda_k$, then
\begin{equation}\label{eq:A2}
\sum_{k=0}^{n-1}\lambda_k(A^*A)\geq\sum_{k=0}^{n-1}\lambda_k^2
\end{equation}
with equality if and only if $A$ is normal.
\end{lemma}

\begin{theorem}[Parseval's identity]Let $f(\lambda)$ be a function that is square-integrable on $[0,2\pi]$, then the sum of the squares of the Fourier coefficients of the function is equal to the integral of the square of the function,
\begin{equation*}
\sum_{-\infty}^{\infty}|t_k|^2=\frac{1}{2\pi}\int_{0}^{2\pi}f(\lambda)^2 d\lambda.
\end{equation*}
\end{theorem}

Consider the truncated Fourier series
\begin{equation}\label{eq:newf}
\hat{f}_n(\lambda)=\sum_{k=-(n-1)}^{n-1}t_ke^{ik\lambda}, \qquad \lambda\in[0,2\pi]
\end{equation}
After some simple analysis and calculation, we know that
\begin{equation*}
\begin{aligned}
\frac{1}{2\pi}\int_0^{2\pi}\!\!\!\!\!\!f(\lambda)\hat{f}_n(\lambda)\ud\lambda=\frac{1}{2\pi}\int_0^{2\pi}\!\!\!\hat{f}^2_n(\lambda)\ud\lambda
=\!\!\!\sum_{k=-(n-1)}^{n-1}\!\!\!\!\!\!|t_k|^2.
\end{aligned}
\end{equation*}
Therefore
\begin{small}
\begin{equation*}
\begin{aligned}
&\frac{1}{2\pi}\int_0^{2\pi}[f(\lambda)-\hat{f}_n(\lambda)]^2\ud\lambda\\
&=\frac{1}{2\pi}\Bigg\{\int_0^{2\pi}\!\!\!f^2(\lambda)\ud\lambda-2\int_0^{2\pi}\!\!\!f(\lambda)\hat{f}_n(\lambda)\ud\lambda+\int_0^{2\pi}\!\!\!\hat{f}_n^2(\lambda)\ud\lambda\Bigg\}\\
&=\frac{1}{2\pi}\int_0^{2\pi}f^2(\lambda)\ud\lambda-\sum_{k=-(n-1)}^{n-1}|t_k|^2.
\end{aligned}
\end{equation*}
\end{small}
Since
\begin{equation*}
\frac{1}{2\pi}\int_0^{2\pi}[f(\lambda)-\hat{f}_n(\lambda)]^2\ud\lambda\geq0
\end{equation*}
it follows that
\begin{equation*}
\sum_{k=-(n-1)}^{n-1}|t_k|^2\leq\frac{1}{2\pi}\int_0^{2\pi}f^2(\lambda)\ud\lambda.
\end{equation*}
This means the sequence $\{t_k\}$ is square summable as $n\to \infty $. Thus given $\epsilon$, there is a single $N$, such that
\begin{equation*}
\begin{aligned}
\sum_{k=-\infty}^{-n}|t_k|^2+\sum_{k=n}^{\infty}|t_k|^2\leq\epsilon \quad \textrm{if}\  n\geq N.
\end{aligned}
\end{equation*}

Because the generating function is a continuous function on $[0,2\pi]$, it is square-integrable. Using Parseval's identity, when $n\geq N$, we have

\begin{equation*}
\begin{aligned}
&\frac{1}{2\pi}\int_0^{2\pi}[f(\lambda)-\hat{f}_n(\lambda)]^2\ud\lambda\\
&=\frac{1}{2\pi}\int_0^{2\pi}f^2(\lambda)\ud\lambda-\sum_{k=-(n-1)}^{n-1}|t_k|^2\\
&=\sum_{-\infty}^{\infty}|t_k|^2-\sum_{k=-(n-1)}^{n-1}|t_k|^2\\
&=\sum_{k=-\infty}^{-n}|t_k|^2+\sum_{k=n}^{\infty}|t_k|^2\\
&\leq\epsilon
\end{aligned}
\end{equation*}

Similar to the proof in \cite{Toe}, we define a circulant matrix $C_n(\hat{f}_n)$ which have top row $(\hat{c}_0,\hat{c}_1 ,\cdots, \hat{c}_{n-1})$, where
\begin{equation}\label{eq:newc}
\hat{c}_k=\frac{1}{n}\sum_{j=0}^{n-1}\hat{f}_n(2\pi j/n)e^{2\pi ijk/n}.
\end{equation}
Since $C_n(f) - C_n(\hat{f_n})$ is a circulant matrix and it is a normal matrix (Corollary \ref{corollary:1}),
we know from  (\ref{eq:Fnorm}) and (\ref{eq:A2}) that
\begin{equation*}
|C_n(f)-C_n(\hat{f_n})|^2=\sum_{k=0}^{n-1}|f(2\pi k/n)-\hat{f}_n(2\pi k/n)|^2
\end{equation*}
and hence for $n\geq N$
\begin{equation}\label{eq:error1}
\begin{aligned}
&\frac{|C_n(f)-C_n(\hat{f_n})|^2}{|T_n(f)|^2}\\
&=\frac{\sum\limits_{k=0}^{n-1}|f(2\pi k/n)-\hat{f}_n(2\pi k/n)|^2}{\sum\limits_{k=-(n-1)}^{n-1}(n-|k|)|t_k|^2}\\
&\leq\frac{\int_0^{2\pi}[f(\lambda)-\hat{f}_n(\lambda)]^2\ud\lambda}{\sum\limits_{k=-(n-1)}^{n-1}(n-|k|)|t_k|^2}\\
&\leq\epsilon
\end{aligned}
\end{equation}
From the result of \cite{Toe},
\begin{equation*}
C_n(\hat{f_n})-T_n(f)=
\left( \begin{array}{ccccc}
0              & t_{n-1}    & t_{n-2}    & \ldots &  t_1   \\
t_{-(n-1)}     & 0          & t_{n-1}    &        &        \\
t_{-(n-2)}     & t_{-(n-1)} & 0          &        &\vdots  \\
\vdots         &            &            & \ddots &        \\
t_{-1}         &            & \ldots     &        &  0     \\
\end{array} \right)
\end{equation*}
therefore
\begin{small}
\begin{equation}\label{eq:error2}
\begin{aligned}
&\frac{|C_n(\hat{f_n})-T_n(f)|^2}{|T_n(f)|^2}\\&=\frac{\sum\limits_{k=-(n-1)}^{n-1}|k||t_k|^2}{\sum\limits_{k=-(n-1)}^{n-1}(n-|k|)|t_k|^2}\\
&=\frac{\sum\limits_{k=-N}^{N}|k||t_k|^2}{\sum\limits_{k=-(n-1)}^{n-1}(n-|k|)|t_k|^2}
+\frac{\sum\limits_{k=-(n-1)}^{-(N+1)}|k||t_k|^2+\sum\limits_{k=N+1}^{n-1}|k||t_k|^2}{\sum\limits_{k=-(n-1)}^{n-1}(n-|k|)|t_k|^2}\\
&\leq\frac{\sum\limits_{k=-N}^{N}|k||t_k|^2}{\sum\limits_{k=-(n-1)}^{n-1}(n-|k|)|t_k|^2}
+\frac{\sum\limits_{k=-(n-1)}^{-(N+1)}|k||t_k|^2+\sum\limits_{k=N+1}^{n-1}|k||t_k|^2}{n|t_0|^2}\\
&\leq\frac{\sum\limits_{k=-N}^{N}|k||t_k|^2}{\sum\limits_{k=-N}^{N}(n-|k|)|t_k|^2}
+\frac{\sum\limits_{k=-\infty}^{-(N+1)}|t_k|^2+\sum\limits_{k=N+1}^{\infty}|t_k|^2}{|t_0|^2}\\
&\leq\frac{N}{n-N}+O(1)\epsilon\\
\end{aligned}
\end{equation}
\end{small}

Since $\epsilon$ is arbitrary,
\begin{equation}\label{eq:inequation}
\begin{aligned}
&\frac{|C_n(f)-T_n(f)|}{|T_n(f)|}\\
&\leq\Bigg(\frac{|C_n(f)-C_n(\hat{f_n})|}{|T_n(f)|}+\frac{|C_n(\hat{f_n})-T_n(f)|}{|T_n(f)|}\Bigg)\\
&\leq\sqrt{\frac{N}{n-N}}
\end{aligned}
\end{equation}
Thus,
\begin{equation}
\lim_{n\rightarrow\infty}\frac{|C_n(f)-T_n(f)|}{|T_n(f)|}=0
\end{equation}

\section{Proof of the Corollary \ref{corollary:2}}
In this appendix, we analyze the magnitude of $\epsilon$, when
\begin{equation*}
|t_k|=
\left\{ \begin{array}{cc}
k^{-p}  &  k\neq0 \\
&\\
1               &  k=0    \\
\end{array} \right.
\end{equation*}
where $p>1$, and $k=0,1,2,\ldots$.

From (\ref{eq:error1}),(\ref{eq:error2}),(\ref{eq:inequation})
\begin{small}
\begin{equation*}
\begin{aligned}
\frac{|C_n(f)-T_n(f)|}{|T_n(f)|}&\leq\frac{|C_n(f)-C_n(\hat{f_n})|+|C_n(\hat{f_n})-T_n(f)|}{|T_n(f)|}\\
                                &\leq\frac{\sum\limits_{k=-\infty}^{-(n-1)}|t_k|^2+\sum\limits_{k=n-1}^{\infty}|t_k|^2+\sum\limits_{k=-(n-1)}^{n-1}|k||t_k|^2}{\sum\limits_{k=-(n-1)}^{n-1}(n-|k|)|t_k|^2}\\
                                &=\frac{2\sum\limits_{k=(n-1)}^{\infty}|t_k|^2+2\sum\limits_{k=0}^{n-1}|k||t_k|^2}{nt_0+2\sum\limits_{k=1}^{n-1}(n-|k|)|t_k|^2}\\
&\leq\frac{\frac{\pi^2}{3}+2+\frac{2}{2^{2p-1}}+\frac{2}{3^{2p-1}}+\cdots\frac{2}{{(n-1)}^{2p-1}}}{nt_0+2(n-1)+(n-2)\frac{2}{2^{2p}}+\cdots+\frac{2}{(n-1)^{2p}}}\\
&=
\left\{ \begin{array}{cc}
O(\textrm{ln}n/n) &  1<p<1.5 \\
                        &          \\
O(1/n)             & 1.5\leq p \\
\end{array} \right.
\end{aligned}
\end{equation*}
\end{small}
where the final two steps use

\begin{equation*}
\lim_{n\rightarrow\infty}\Big(1+\frac{1}{2}+\frac{1}{3}+\cdots+\frac{1}{n}-\textrm{ln}n\Big)=\gamma
\end{equation*}

\begin{equation*}
\sum_{n=1}^{\infty}\frac{1}{n^2}=1+\frac{1}{2^2}+\frac{1}{3^2}+\cdots=\frac{\pi^2}{6}
\end{equation*}
and $\gamma$ is Euler-Mascheroni constant. In view of (\ref{eq:ferror}), for the well-conditioned Toeplitz matrices, the magnitude of final error is
\begin{equation*}
O(\frac{\textrm{ln}n\textrm{log}n}{n}).
\end{equation*}

\section{ Proof of the Corollary \ref{corollary:3}}

The corollary is proven from following theorem:

\begin{theorem}[\cite{square}]Let $T_n$ be a family of Toeplitz Hermitian matrices associated with the sequence $\{t_k\}$, and $F(\lambda)$ be the discrete-time Fourier transform (DTFT) of $\{t_k\}$. If $|F(\lambda)|\neq0 $ for $\lambda\in[0,2\pi]$ and $\sum^{\infty}_{k=-\infty}|kt_k|<\infty$, for a vector with the form
\begin{equation*}
  x=(0,\ldots,0,x_{-L},\ldots,x_0,\ldots,x_L,0,\ldots,0),
\end{equation*}
 the quadratic form is bounded by
\begin{equation*}
  \frac{\|(T_n^{-1}-C_n^{-1})x\|}{\|x\|}\leq O(1/\sqrt{n}).
\end{equation*}
\end{theorem}
It is easy to verify that the conditions of the theorem are met when the generating function is a strictly positive continuous real-valued function. Besides, because of $\lambda_{max}(T_n)\leq 1$, that
\begin{equation*}
 \frac{\|(T_n^{-1}-C_n^{-1})x\|}{\|T_n^{-1} x\|} \leq\frac{\|(T_n^{-1}-C_n^{-1})x\|}{\|x\|}\leq O(1/\sqrt{n}).
\end{equation*}
Thus, the magnitude of final error is $O(1/\sqrt{n})$.

\vspace*{15pt}

\section*{Acknowledgements}
This work is supported by National Natural Science Foundation of China (Grant Nos. 61672110, 61671082).

\bibliography{Toeplitz}

\end{document}